\documentclass{ws-ijmpd}

\begin{document}

\markboth{John Swain}
{Entropy and Area in Loop Quantum Gravity}

%%%%%%%%%%%%%%%%%%%%% Publisher's Area please ignore %%%%%%%%%%%%%%%
%
\catchline{}{}{}{}{}
%
%%%%%%%%%%%%%%%%%%%%%%%%%%%%%%%%%%%%%%%%%%%%%%%%%%%%%%%%%%%%%%%%%%%%

\title{ENTROPY AND AREA IN LOOP QUANTUM GRAVITY}
\author{JOHN SWAIN}

\address{Department of Physics, Northeastern University, 110 Forsyth Street\\
Boston, MA 02115\\
Canada\\
john.swain@cern.ch}

\maketitle

\begin{history}
\received{Day Month Year}
\revised{Day Month Year}
\comby{Managing Editor}
\end{history}

\begin{abstract}
Black hole thermodynamics suggests that the maximum entropy 
that can be contained in a region of space 
is proportional to the area enclosing it rather
than its volume. I argue that this
follows naturally from loop quantum gravity and a result of Kolmogorov and Bardzin'
on the the realizability of networks in three dimensions. This represents
an alternative to other approaches in which some sort of correlation
between field configurations helps limit the degrees
of freedom within a region. It also provides an approach to thinking
about black hole entropy in terms of states inside rather than 
on its surface.
Intuitively, a spin network complicated enough to imbue a region
with volume only lets that volume grow as quickly as the area bounding
it.
\end{abstract}


\section{General Appearance}

In quantum field theory without gravity and without any fundamental length
scale, the entropy which can be contained in a region $R$ of volume
$V(R)$ is proportional to volume times a formally infinite factor.

To see this in a simple model, let us follow Susskind \cite{Susskind}\footnote{This is also
a good reference to the ideas of holography where the idea is that our 3 dimensional
world is somehow, more fundamentally, 2 dimensional.}, and think of the region
$R$ as a lattice of spins separated by some minimum length ({\it i.e.} the Planck
length). 
The number $N(V)$ of orthogonal (distinct, independent)  degrees of freedom then
grows as $N(V) = 2^n$ where $n$ is the number of lattices sites in $R$. $n$ is clearly
proportional to the volume $V(R)$. The maximum entropy $S_{max}$ which can be contained in
$R$ is then the logarithm of this, so we have $S_{max} = n \log(2) \propto V$. In the
absence of a minimum length, $n\rightarrow \infty$ and the entropy is infinite, so 
gravity already helps by introducing a cutoff. However, a problem still remains.

Arguments from black hole thermodynamics suggest that the maximum
entropy that can be contained in $R$ should be proportional not to its volume $V(R)$,
but rather the area of its boundary $A(\partial R)$. Were this not the case, one
could violate the second law of thermodynamics, and the argument is well-known:
Many states contributing to the entropy described above for a lattice of spins
correspond to so much energy in $V(R)$ that a black hole would form. One can pick
a configuration close to being a black hole, add a little matter, and form 
a lower entropy black hole, in violation of the second law.

How then can one resolve this problem? Detailed calculations in string theory certainly
have had some successes\cite{Horowitz}, but string theory is a very complicated structure
containing much besides gravity, and I want to concentrate here on what gravity alone can do.
Note that loop quantum gravity \cite{Rovelli} also provides a means of understanding black hole
entropy in terms of surface degrees of freedom but a horizon is needed.

Gravity, at least intuitively, provides a fundamental
length and thus a means for regularizing infinities in energies. Indeed, this is the case
for finiteness of ADM mass for a charged sphere of radius tending to zero in
Einstein-Maxwell theory (as well as in the simple Newton-Coulomb analog
described by Ashtekar in \cite{Ashtekar}). Could gravity also provide a way to not
only {\em regularize an infinite entropy},
but also {\em change the scaling behaviour for entropy} from
following volume to following area?

A very simple argument by Yurtsever\cite{Yurtsever} which comes close in spirit to what I 
propose, is that a maximum entropy proportional to area comes about naturally
from the assumption that there is a constraint on the {\em total energy} of the Fock space states
which are allowed to contribute. Essentially what this boils down to is that some
agent, which we could take to be gravity, steps in to ensure that the energy in $R$ cannot
grow without limit and constrains not one mode or another singly, but the entire set of modes that
can be allowed.

Here we start to see a hint at what a fundamental underlying mechanism might be: one
cannot get the correct entropy if one thinks in terms of modes of free fields which do not
interact with one another, and in some sense some of these modes must get ``in the
way'' of each other in the sense that they cannot grow in energy without limit inside
some finite volume since their maximum sum is bounded.

Here I argue that within loop quantum gravity (LQG)\cite{Rovelli}there is also a natural regularization
of entropy, and that this regularization constrains the maximum entropy of
a region of space to grow no faster than its area. This is perhaps a surprising result
since one might imagine that even with a discrete spacetime of the type with
which this essay started (with a cutoff provided by gravity) one would still find a number
of lattice points inside a macroscopic volume which grows like that volume. 
The argument is very generic and relies only thinking about what a spin network 
means in terms of geometric observables. 

Recall that a spin network is a graph
whose edges are labelled with $SU(2)$ representations $j$ and whose vertices are 
intertwiners. For a region of space $R$ bounded by $\partial R$, the area 
spectrum is discrete and 
is given by a sum of contributions proportional to $\sqrt{j(j+1)}$ from each edge with label $j$
that punctures it. The spectrum of the volume is discrete and given by a more
complex (and still not fully understood\cite{vol}) expression with contributions from each 
vertex with more than 3 edges. Here the details are unimportant and all that matters is
the fact that the spectrum is discrete so that $V(R)$ is bounded by
some constant times the number of vertices inside -- that is, the vertices basically represent
little elements of volume. This is all we'll need of loop quantum gravity.

Now let us appeal to a little-known theorem\footnote{I learned about it first from
V. I. Arnold's obituary of Kolmogorov in Physics Today, October 1989 where he
explains that Kolmogorov was motivated by ideas from the physiology 
of the brain.} of Kolmogorov and Bardzin'\cite{Kolmogorov}.
Originally stated with electronic logic circuits in mind, the idea is simple.
Consider some cubical region of space $R$ ({\em i.e.} a box) and try to build a computer in it. 
If you try to pack a circuit into it made of elements
of some finite size connected by wires of some finite thickness, with some minimum distance
between the elements, then provided the
circuit is complicated enough, the maximum number of elements you can put inside grows
not like the volume of $R$ but of the area that encloses it. In intuitive terms,
the finite-thickness wires of a sufficiently complicated circuit get in the way so much
that, as you put a large number of elements in, packing as closely as
you can, you only get to put in a number of elements that grows
as the area of the box's surface. Essentially you lose a dimension to the wires, or to
the need for elements to ``communicate with each other'', or be ``linked'' or ``connected''.
The term ``sufficiently complicated'' basically removes degenerate cases like a long
circuit made of elements all connected by wires in one long line.

Now consider a spin network. To each edge you must assign a cross sectional area
since any surface gets its geometrical area precisely from its being
crossed by spin network edges. Similarly to each vertex you must assign a volume
since any region of space gets its geometrical volume precisely from the presence
of vertices. Finally, the clause about ``sufficiently complicated'' 
in the previous paragraph, is taken care of by the need to have quadrivalent
vertices.

In other words, it would seem that
any region of spacetime (horizon present or not) bounded by
some area cannot contain a spin network whose number of nodes (chunks of volume)
grows asymptotically faster than the area. If one has additional fields, one
expects the same argument to hold - gravity regularizes the field theoretic divergences
and also changes the nature of the support for the fields (chunks of volume) so that asymptotically,
in a sense, the volume over which integrations are performed
only grows like area.

A point of clarification should be made here which did not appear so explicitly in the first
version of this paper. One might imagine running into trouble by arguing as follows:
Consider a spin network which has given volume and area to a region of space as described
above. Following Perez\cite{Perez} one can now imagine choosing some internal part of
the spin network and adding
one quadrivalent vertex with edges connecting to other internal edges. In fact, this
could even be done starting with the simplest nontrivial network: one quadrivalent
vertex corresponding to volume with the 4 edges coming out and puncturing an enclosing
surface to imbue it with area. This will
increase the volume inside the region without changing its area, since the quadrivalent
node will carry volume but add no new edges which carry area for the region. In other
words, it might seem that a given fixed area can enclose an arbitrarily large volume!

What is being argued here is precisely that this may not be the case when one thinks about
spin networks physically. In other words,
instead of writing down a spin network in the usual way, thinking of the edges as 1-dimensional
and the nodes as 0-dimensional, one should take the physical interpretation seriously and
``thicken'' the edges so that they have 2-dimensional cross-sections and the nodes so
that they are small 3-dimensional balls. The actual sizes are unimportant as long
as there is some minimum, which we already know from the quantization of area and
volume in loop quantum gravity. 

Now  while one might write down a spin network in the usual way (not thinking of the edges and
vertices as essentially thickened as described in the previous paragraph) and get, for some area,
an arbitrarily large volume, this idea does not fit with the Kolmogorov-Bardzin' result.
It suggests that there must be some additional, perhaps
dynamical, feature that constrains what spin networks one would have to use to specify
spacetime geometries. I argue that one can already see at least part of what that
must be from the results presented here in terms of replacing a spin-network by a
physically-motivated thickened graph.

While naively gravity provides a cutoff at short distances, it also ``gets in
the way of itself'' in such a way as to reduce the effective dimensionality 
of spacetime so that asymptotically as one tries to include more and more degrees of freedom
``volume'' (as counted by nodes of a spin network inside) scales like ``area''! 
It is intriguing to think of this as a loop quantum gravity variant of an
old idea of Crane and Smolin\cite{CraneSmolin}
on ``spacetime foam as a universal regulator'' where regularization comes about
due to a small decrease in the effective dimension of spacetime. Here something 
similar happens in that the microstructure of space is modified in LQG, but now
the dimension of space is not reduced by a small amount, but all the way from 
3 to 2 as distances become small!

\section{Acknowledgements}

I would like to thank the US National Science Foundation 
for continued support and my colleagues
in various parts of the world who have shared discussions with me. In
particular, I would like to thank Lee Smolin for
a few discussions at one time or another, and Carlo Rovelli for the invitation
to ``Non Perturbative Quantum Gravity: Loops and Spin Foams''
in Marseilles in 2004 where I started to think more about this problem.
I would also like to thank Prof. Ya. M. Bardzin' for sending me
a copy of his paper in Russian and Yuri Musienko for his translation of
that paper to English. Finally, I would like to thank Alejandro Perez for
thoughtful comments on the first draft of this paper, and in particular
for indicating the importance that must be given to the possible role of dynamics
in an ultimate explanation of the connections between area and volume in
loop quantum gravity.

\section{References}

%\begin{thebibliography}{000} %for 3 digits
%\begin{thebibliography}{00}  %for 2 digits

\end{document}